\documentclass[10pt]{iopart}
\expandafter\let\csname equation*\endcsname\relax
\expandafter\let\csname endequation*\endcsname\relax

\usepackage{amsmath}
\usepackage{graphicx}
\usepackage{lipsum}
\usepackage{bm}
\usepackage{tabu}
\usepackage[utf8]{inputenc}
\usepackage[T1]{fontenc}
\usepackage{mathptmx}
\usepackage{color}
\usepackage{xcolor}
\usepackage{cite}
\usepackage{iopams}
\usepackage{cancel}

\newcommand{\vect}[1]{\ensuremath{\mathbf #1}}

\newcommand{\dd}[3]{\ensuremath{\frac{d^\text{#3} #1}{d #2^\text{#3}}}}

\newcommand{\MGe}{\text{MG}^e}
\newcommand{\MGo}{\text{MG}^o}

\newcommand{\GB}{\text{GB}}

\newcommand{\Je}{\text{Je}}
\newcommand{\Jo}{\text{Jo}}
\newcommand{\ce}{\text{ce}}
\newcommand{\se}{\text{se}}

\newcommand{\ket}[1]{\ensuremath{\left|#1\right\rangle}}
\newcommand{\braket}[2]{\ensuremath{\left\langle #1 | #2 \right\rangle}}

\begin{document}
\title[Violation of Bell's inequality for Mathieu-Gauss vector modes]{Violation of Bell's inequality for Mathieu-Gauss vector modes}

\author{Edgar Medina-Segura$^1$, Francisco I. Mecillas-Hern\'andez$^1$, Thomas Konrad$^2$, Carmelo Rosales-Guzm\'an$^{1,\dag}$ and Benjamin Perez-Garcia$^{3,*}$}

\address{$^1$ Centro de Investigaciones en Óptica, A.C., Loma del Bosque 115, Colonia Lomas del campestre, C.P. 37150 León, Guanajuato, Mexico.}
\address{$^2$ School of Chemistry and Physics, University of Kwazulu-Natal, Private Bag X54001, Durban 4000, South Africa.}
\address{$^3$ Photonics and Mathematical Optics Group, Tecnologico de Monterrey, Monterrey 64849, Mexico.}
\ead{$^{*}$b.pegar@tec.mx, $^{\dag}$carmelorosalesg@cio.mx}

\begin{abstract}
Vector beams display varying polarisation over planes transversal to their direction of propagation. The variation of polarisation implies that the electric field cannot be expressed as a product of a spatial mode and its polarisation. This non-separability has been analysed for particular vector beams in terms of non--quantum entanglement between the spatial and the polarisation-degrees of freedom, and equivalently, with respect to the degree of polarisation of light. Here we demonstrate theoretically and experimentally that Mathieu-Gauss vector modes violate a Bell-like inequality known as the Clauser-Horn-Shimony-Holt-Bell (CHSH-Bell) inequality. This demonstration provides new insights into that fact that a more general class of vector modes with elliptical symmetry also violate Bell inequalities.
\end{abstract}

\ioptwocol
\maketitle
\section{Introduction}
Entanglement constitutes one of the cornerstones of quantum mechanics that gives rise to exceptional properties. For example, entangled particles seem to be able to communicate instanteneously with each other, which would violate the principles of relativity. This paradoxical behaviour, dismissively referred as “spooky actions at a distance” by Einstein \cite{EPR_Paradox}, is generated by the superposition of different combinations of states of two or more systems, which according to  classical physics could only occur if the choice of observable measured on one system would influence the outcome of  measurements on the other system(s), regardless of their separation -- a property known as nonlocality. The concepts of entanglement and nonlocality are closely related, challenging our understanding of physical reality. In 1964, Bell proposed a test for all theories that are local (i.e. no instantaneous communication between remote systems)  and realistic (systems have detected properties before measurement) by deriving an inequality for correlations based on these assumptions \cite{Bell1964}. It has then become a standard procedure to test the ``quantumness'' of a system through the violation of Bell’s inequality \cite{Bell1966}, which for optical systems usually takes the form of Clauser-Horne-Shimony-Holt-Bell (CHSH-Bell) inequality \cite{CHSH1969}. However, entanglement occurs also between different degrees of freedom of the same quantum system if its state consists of a superposition of different combinations of states in the two degrees of freedom, which cannot be factorised, i.e. written as a product of the state of the first degree of freedom and the state of the second degree of freedom. For example an electromagnetic wave with elementary excitation (a single photon) can be written as a vector mode of the form  ${\bf u}({\bf r}) = \cos(\alpha/2) u^+({\bf r}) \hat{e}_R + \sin(\alpha/2) u^-({\bf r}) \hat{e}_L$, which is a superposition of a certain mode $u^+({\bf r})$ with right-circular ${\bf \hat{e}}_R$ polarisation and a different mode, $u^+({\bf r})$,  with  left-circular polarisation ${\bf \hat{e}}_L$. Such a vector mode differs from a scalar light field, which can be represented by a single product of a  spatial mode and the polarisation state $u({\bf r}) {\bf \hat{e}}$ and corresponds to an entangled state that violates a Bell inequality, as opposed to a single photon of a scalar light field with separable spatial mode and polarisation.

A major contribution to the understanding of entanglement was the realisation that it occurs in similar form in classical field theories, such as electromagnetism, and there leads to features formerly thought to be reserved to quantum systems.  For example, vector beams in classical optics (corresponding in quantum optics to coherent states with classical properties)  can also violate Bell inequalities \cite{Ndagano2016,forbes2019classically,Selyem2019,Qian2011,Aiello2015,konrad2019,YaoLi2020}, or they can be used to show that not all optical operations (Mueller matrices) can be realised in the lab, but only those that are completely positive \cite{Simon2010}.

The entanglement between different degrees of freedom of the same quantum system or electromagnetic wave can of course not be used to refute local realistic theories via the violation of a Bell inequality, - optics is a local realistic theory after all. The reason for this is, that the measurements on different degrees of freedom of the same system cannot be spatially separated and, therefore, do not allow to exclude the influence of the measurement of one variable on the outcomes the measurement of another. On the other hand, strong violation of a Bell inequality by classical light can indicate strong entanglement of classical degrees of freedom, and vice versa, i.e. entanglement measures such as Concurrence can be used to quantify the ``vectorness'' of a light beam \cite{McLaren2015, Otte2018Bell}.



%
In this manuscript we study the CHSH-Bell inequalities for Helical-Mathieu Gauss Vector modes (HMGVM), which are characterised by their elliptical spatial profile controlled by the ellipticity parameter $\epsilon\in[0, 1]$, which changes the shape of the mode, from circular ($\epsilon=0$) to elliptical ($0<\epsilon<1$). Here, we derive a general expression for the CHSH-Bell parameter {\it S} that depends on $\epsilon$, which for $\epsilon=0$ reduces to the well-know expression derived for Laguerre-Gaussian modes \cite{McLaren2015}. The manuscript is organised as follows, first in section \ref{MathieuVEctorModes} we briefly describe the HMGVM, as first introduced in \cite{Rosales2021}, but in this case we use Dirac's notation to highlight their similarity with entangled states. Afterwards, in section  \ref{CHSH} we derive a theoretical expression for the CHSH-Bell parameter. In section \ref{Experiment} we describe the experimental techniques behind the generation of HMGVM as well as the technique to measure CHSH-Bell inequalities. Finally, we conclude this manuscript with a discussion section.


\section{Mathematical framework}
\subsection{Helical Mathieu-Gauss vector beams}\label{MathieuVEctorModes}
\begin{figure}[tb]
    \centering
    \includegraphics[width=0.49\textwidth]{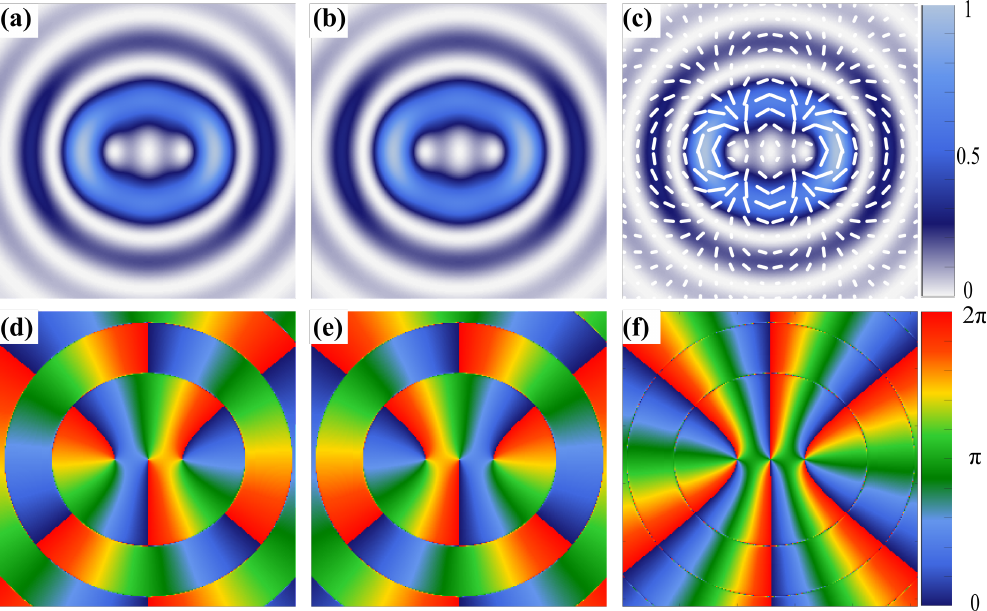}
    \caption{Theoretical intensity (first row) and phase (second row) distribution of the circular polarisation components and Stokes field ($S = S_1 + iS_2$), for $\vect{u}_3(\vect{r};q)$.  (a) Intensity and (d) phase of the right--handed circular polarisation component. (b) Intensity and (e) phase of the left--handed circular polarisation component.  (c) Depicts the intensity of the Stokes field overlapped with the non--homogeneous distribution of polarisation states (white lines).  (d)  Shows the phase of the Stokes field.}
    \label{concept}
\end{figure}

Helical Mathieu-Gauss Vector (HMGV) beams are generated as a non-separable superposition of the polarisation degree of freedom and the Helical Mathieu-Gauss (HMG) beams, which are expressed as \cite{Rosales2021}
\begin{equation}
    {\bf u}_{m}(\vect{r};q) = \cos\left(\frac{\alpha}{2}\right) u_m^+(\vect{r};q)\hat{\bf e}_R+\sin\left(\frac{\alpha}{2}\right)\exp(i\beta)u_m^-(\vect{r};q)\:\hat{\bf e}_L,
    \label{eq:HMVB}
\end{equation}
where $\vect{r}=(r,\varphi,z)=(\xi,\eta,z)$ defines the polar-- and elliptical--cylindrical coordinates, respectively; $\alpha\in[0,\pi/2]$ is a weighting coefficient that determines the contribution of each polarisation component and the exponential term $\exp(i\beta)$ ($\beta\in [0, \pi]$) is an inter-modal phase between both polarisation components. The terms $\hat{\bf e}_R$ and $\hat{\bf e}_L$ represent the right and left handed unitary vectors of the circular polarisation basis, respectively.  The basis in the spatial degree of freedom is defined through the modes $u_m^{\pm}(\vect{r};q) = \MGe_m(\vect{r};q) \pm i \MGo_m(\vect{r};q)$ \cite{Gutierrez-Vega2005} with
\begin{align}
    \text{MG}_m^{e}(\vect{r};q) &=\exp\left({-\frac{ik_t^2}{2k}\frac{z}{\mu}}\right)\GB({\bf r})\Je_m(\tilde{\xi},q)\ce_m(\tilde{\eta},q),\\
    \text{MG}_m^{o}(\vect{r};q) &=\exp\left({-\frac{ik_t^2}{2k}\frac{z}{\mu}}\right)\GB({\bf r})\Jo_m(\tilde{\xi},q)\se_m(\tilde{\eta},q),
\end{align}
where $z$ is the propagation distance, and the complex elliptic coordinates $(\tilde{\xi},\tilde{\eta})$ are determined by
\begin{align}
    x &= f_0 (1+iz/z_R)\cosh{\tilde{\xi}} \cos{\tilde{\eta}},\\
    y &= f_0 (1+iz/z_R)\sinh{\tilde{\xi}} \sin{\tilde{\eta}}.
\end{align}
The parameter $f_0$ is the semifocal separation of the elliptical coordinates at $z=0$, which relates to the  major and  minor axis $a$ and $b$, respectively, as $f^2=a^2-b^2=a\varepsilon$ and $q=(f k_t/2)^2$ is a dimensionless parameter.  In addition, the function $\GB({\bf r})$ is the fundamental Gaussian beam defined as
\begin{equation}\label{MGmoes}
    \GB({\bf r})=\exp\left({-\frac{r^2}{\mu\omega_0^2}}\right)\frac{\exp(ikz)}{\mu},\\
\end{equation}
where $\mu=\mu(z)=1+iz/z_R$, $z_R=k\omega_0^2/2$ is the Rayleigh range of a Gaussian beam with waist radius $\omega_0$. The parameters $k_z$ and $k_t$ are the longitudinal and transverse components of the wave vector $\bf k$, whose magnitude $k=2\pi/\lambda$ satisfies the relation $k^2=k_t^2+k_z^2$. Further, the functions $\Je_m$ and $\Jo_m$, are the  $m$th-order even and odd radial Mathieu functions, respectively, and $\ce_m$ and $\se_m$ are the even and odd  $m$th-order angular Mathieu functions \cite{NIST2010}. By way of example, Fig. \ref{concept} schematically illustrates the intensity and phase profiles of the two orthogonal scalar Mathieu-Gauss beams, which are recombined in a nonseparable fashion to produce its vector counterpart. Here, the HMG modes $u_m^{\pm}(\vect{r};q)$, whose transverse intensity and phase profile are shown in Figs. \ref{concept}(a), \ref{concept}(d) and \ref{concept}(b) and \ref{concept}(e), respectively, are coaxialy superposed in a non-separable fashion to generate the vector mode shown in Fig.\ \ref{concept} (c).  The phase of the complex Stokes field $S=S_1 + iS_2$ is shown in Fig. \ref{concept} (f). The specific parameters for this example are $k_t=4$ mm$^{-1}$ , $e=0.9$, $a=1.5$ and $m=3$.  

Importantly, a careful use of Dirac's notation, allows us to express Eq.\ \ref{eq:HMVB} as
\begin{align}\label{eq:vectorstate}
    \ket{\psi_m}_q = \cos\left(\frac{\alpha}{2}\right) \ket{m_q^+}\ket{R} + \sin\left(\frac{\alpha}{2}\right)\exp(i\beta)\ket{m_q^-}\ket{L},
\end{align}
\noindent
where the kets $\ket{m_q^\pm} = (\ket{m_q^e} \pm i\ket{m_q^o})/\sqrt{2}$ represent the unit vectors in an infinite dimensional Hilbert space $\mathcal{H}_\infty$, describing the spatial mode on a transverse plane at a fixed $z=z_0$.  On the other hand, $\ket{R}$ and $\ket{L}$ represent the right-- and left-- handed polarization basis (two--dimensional Hilbert space).  In addition
\begin{align}
    \braket{\tilde{\xi},\tilde{\eta}}{m_q^+} &= u_m^+(\tilde{\xi},\tilde{\eta},z_0;q),\\
    \braket{\tilde{\xi},\tilde{\eta}}{m_q^-} &= u_m^-(\tilde{\xi},\tilde{\eta},z_0;q),\\
    \braket{\tilde{\xi},\tilde{\eta}}{m_q^e} &= \MGe(\tilde{\xi},\tilde{\eta},z_0;q),\\
    \braket{\tilde{\xi},\tilde{\eta}}{m_q^o} &= \MGo(\tilde{\xi},\tilde{\eta},z_0;q),
\end{align}
where the following orthogonality rules are in place:  $\braket{{m'}_q^{\sigma'}}{m_q^{\sigma}}=\delta_{m,m'}\delta_{\sigma,\sigma'}$ with $\sigma, \sigma' = \{e,o\}$; and the inner products
\begin{align}
    \braket{m_{q'}^{e}}{m_q^{o}} &= 0,\\
    \braket{m_{q'}^{e}}{m_q^{e}} &=
    \begin{cases}
        \sum_{j=0}^\infty A_{2j}^{2n}(q') A_{2j}^{2n}(q), \text{for $m=2n$},\\
        \sum_{j=0}^\infty A_{2j+1}^{2n+1}(q') A_{2j+1}^{2n+1}(q), \text{for $m=2n+1$},\\
    \end{cases}\\
    \braket{m_{q'}^{o}}{m_q^{o}} &=
    \begin{cases}
        \sum_{j=0}^\infty B_{2j+1}^{2n+1}(q') B_{2j+1}^{2n+1}(q), \text{for $m=2n+1$},\\
        \sum_{j=0}^\infty B_{2j+2}^{2n+2}(q') B_{2j+2}^{2n+2}(q), \text{for $m=2n+2$},\\
    \end{cases}
\end{align}
where $A_j^m(q)$ and $B_j^m(q)$ are the coefficients of the Fourier expansion, corresponding to the angular even and odd Mathieu functions, respectively (see \ref{sec:appA}).

\subsection{Bell inequalities}
\label{CHSH}
\begin{figure}[h]
    \centering
    \includegraphics[width=0.49\textwidth]{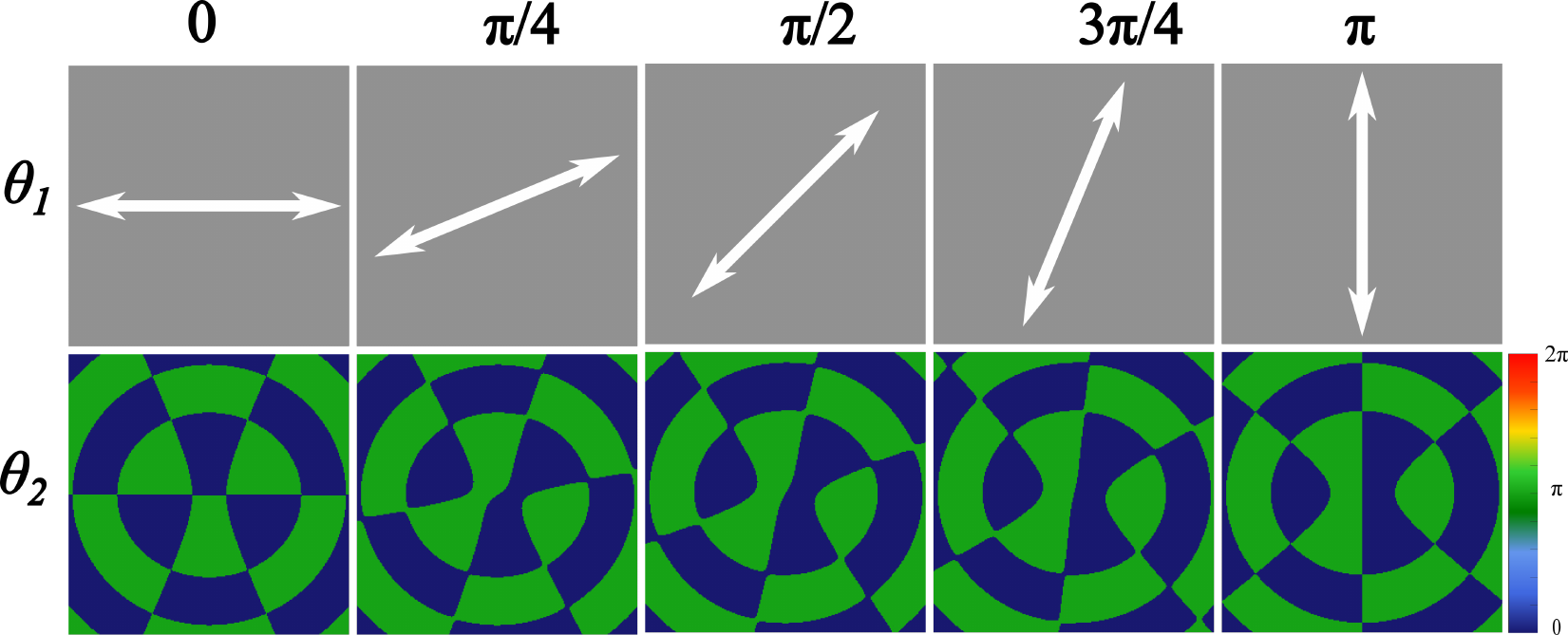}
    \caption{Schematic representation of the projection onto the spatial (top panels) and polarisation (bottom panels) DoFs.}
    \label{projection}
\end{figure}

As explained before, the helical Mathieu-Gauss vector modes are constructed as a non--separable superposition of the spatial and polarisation degrees of freedom, giving rise to classically-entangled states. Even though a formal demonstration that such beams are entangled in the classical sense has not been provided yet, it can be demonstrated through the Clauser-Horne-Shimony-Holt (CHSH) inequality, which is one of the most commonly used Bell-like inequality for optical systems \cite{toninelli2019concepts}. According to \cite{McLaren2015}, the CHSH-Bell parameter $S$ is defined as
\begin{equation}
    S=E(\theta_1,\theta_2)-E(\theta_1,\theta_2')+E(\theta_1',\theta_2)+E(\theta_1',\theta_2').
    \label{BellParameter}
\end{equation}
\noindent
which, for any classical theory based on hidden variables the Bell parameter has an upper bound $|S|\leq2$. In the context of non-separable states of light, $\theta_1$ and $\theta_2$ are the orientation angles of the half-wave plate and the encoded hologram, respectively, as schematically shown in Fig. \ref{projection} for the specific cases $\theta_1=0, \pi/4, \pi/2, 3\pi/4, \pi$ and $\theta_2=0, \pi/4, \pi/2, 3\pi/4, \pi$.  

Further, and also in the context of non-separable classical states  $E(\theta_1,\theta_2)$ is calculated from far-field intensity measurements along the optical axis as
\begin{equation}
\begin{split}
&E(\theta_1,\theta_2)=\\ &\frac{I(\theta_1,\theta_2)+I(\theta_1^{\perp},\theta_2^{\perp})-I(\theta_1^{\perp},\theta_2)-I(\theta_1,\theta_2^{\perp})}{I(\theta_1,\theta_2)+I(\theta_1^{\perp},\theta_2^{\perp})+I(\theta_1^{\perp},\theta_2)+I(\theta_1,\theta_2^{\perp})},
\end{split}
\label{eq:E}
\end{equation}
where $\theta_i^\perp=\theta_i+\pi/2$. Finally, the projection angles of Eq. \ref{BellParameter} are not arbitrary but restricted to the condition
\begin{equation}
    \theta\equiv\theta_2-\theta_1=\theta_2'+\theta_1'=-\theta_2-\theta_1'.
\end{equation}

To measure the intensity $I$, we consider the ``classically entangled'' state as described by Eq. \ref{eq:vectorstate}.  For clarity, let us remind the following inner products
\begin{align}\nonumber
    \braket{m_{q'}^e}{m_{q}^e} &\equiv \frac{1}{\pi}\int_0^{2\pi} \ce_m (\eta,q') \ce_m (\eta,q) d\eta,\\
    &=\begin{cases}
         \sum_{j=0}^{\infty} A_{2j}^{2n}(q') A_{2j}^{2n}(q), \text{ for $m=2n$},\\
         \sum_{j=0}^{\infty} A_{2j+1}^{2n+1}(q') A_{2j+1}^{2n+1}(q), \text{ for $m=2n+1$},
    \end{cases}
\end{align}
\begin{align}\nonumber
    \braket{m_{q'}^o}{m_{q}^o} &\equiv \frac{1}{\pi}\int_0^{2\pi} \se_m (\eta,q') \se_m (\eta,q) d\eta,\\
    &=\begin{cases}
         \sum_{j=0}^{\infty} B_{2j+1}^{2n+1}(q') B_{2j+1}^{2n+1}(q), \text{ for $m=2n+1$},\\
         \sum_{j=0}^{\infty} B_{2j+2}^{2n+2}(q') B_{2j+2}^{2n+2}(q), \text{ for $m=2n+2$},
    \end{cases}
\end{align}
where we have used Eqs.\ \ref{eq:innerce} and \ref{eq:innerse}.

Then, we calculate the intensity associated to Eq.\ \ref{eq:vectorstate} by projecting the polarization degree of freedom onto $\ket{R}+\exp(2i\theta_1)\ket{L}$, and the spatial degree of freedom on $\ket{m^+_{q'}} + \exp(2i\theta_2)\ket{m^-_{q'}}$, which after some algebra leads to
\begin{align}
    I(\theta_1,\theta_2) = &\frac{1}{4} Q(m,q,q') [1+\sin2\alpha\cos(2\theta_1 + 2\theta_2 + \beta)],
    \label{eq:Intensity}
\end{align}
where
\begin{align}\nonumber
    Q(m,q,q') = \bigg|\frac{1}{2}\sum_j [A_j^m(q') A_j^m(q) + B_j^m(q') B_j^m(q)]\bigg|^2.
\end{align}
Here, it is worth highlighting that the case $q=q'$ leads to $Q(m,q,q)=1$ and therefore the intensity $I(\theta_1,\theta_2)$ reduces to the well--known case reported in \cite{McLaren2015}. The Bell parameter $S$ reaches a maximum value $2\sqrt2$ when $\theta=\pi/8$, where we obtain a maximally non-separable state.  Crucially, the value of $S$ does not depend neither on $q$ nor $q'$, this can be seen by plugging Eq.\ \ref{eq:Intensity} in Eq.\ \ref{eq:E}, and realise that $Q(m,q,q')$ appears in both, the numerator and the denominator, and cancels out from this equation. This is a crucial result that will  be demonstrated experimentally  in the next section. As a final comment, it is common to use the angles $\theta_1=0$, $\theta_1'=-\pi/4$, $\theta_2=\pi/8$ and $\theta_2'=3\pi/8$ to violate the CHSH inequality and these are precisely the angles we will use in the experiments described in the next section.


\section{Experimental results}
\label{Experiment}

\subsection{Experimental setup}
\begin{figure}[tb]
    \centering
    \includegraphics[width=0.49\textwidth]{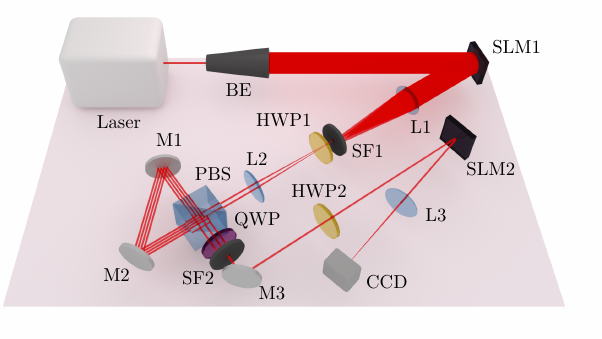}
    \caption{Schematic representation of the optical setup implemented to generate helical Mathieu vector beams and perform projection measurements.  BE: Beam Expander; L1--L3: Lenses; PBS: Polarizing Beam Splitter, SLM: Spatial Light Modulator; HWP1 and HWP2: Half-Wave plates; QWP: Quarter-Wave Plate; M1--M3: Mirrors; SF1 and SF2: Spatial Filters; CCD: Charge Coupled Device Camera}
    \label{setup}
\end{figure}

In order to corroborate experimentally our theoretical results, we implemented a highly stable optical setup to generate arbitrary vector beams, which is schematically illustrated in Fig. \ref{setup} and detailed in \cite{Perez-Garcia2017}. The setup starts with a horizontally polarised HeNe laser beam ($\lambda=$ 632.8 nm), collimated and expanded to cover from edge to edge the liquid crystal screen of a Spatial Light Modulator (SLM1), which in our case is the Pluto LCOS phase only from  Holoeye with a spatial resolution of 1920$\times$1080 pixels and a pixel size of 8 $\mu$m. The screen of the SLM is divided in two halves, each of which is addressed with a digital hologram for the independent generation and manipulations of an optical field. Each holograms is generated using complex amplitude modulation, as detailed in \cite{SPIEbook}, and superimposed with a linear blazed grating to separate all diffraction orders and spatially filter the first diffraction order using a telescope system formed by lenses L1 and L2 ($f=500$ mm) and the spatial filter SF1. The beams emerge from the SLM with a horizontal polarisation state which is then rotated to diagonal using a Half-Wave Plate (HWP1).  Afterwards, they are redirected to a common--path triangular interferometer of the Sagnac type, which starts with a Polarising Beam Splitter (PBS) and incorporates two mirrors. In this way, each beam is separated by the PBS into its horizontal and vertical components travelling inside the interferometer along opposite directions and after a round trip all four beams exit from the port adjacent to the input one. The mirrors of the interferometer are then adjusted to co-axially recombine the beam with horizontal polarisation that emerges from one half of the SLM with the beam with vertical polarisation that emerges from the other half, ensuring in this way orthogonality in both, the polarisation and spatial Degrees of Freedom (DOFs). The generated Helical Mathieu vector beam is then sent to our analysis stage which consist on a Half-Wave plate (HWP2), a second spatial light modulator (SLM2) and a Charge-Coupled Device (CCD) camera. 

\begin{figure}[t]
    \centering
    \includegraphics[width=0.49\textwidth]{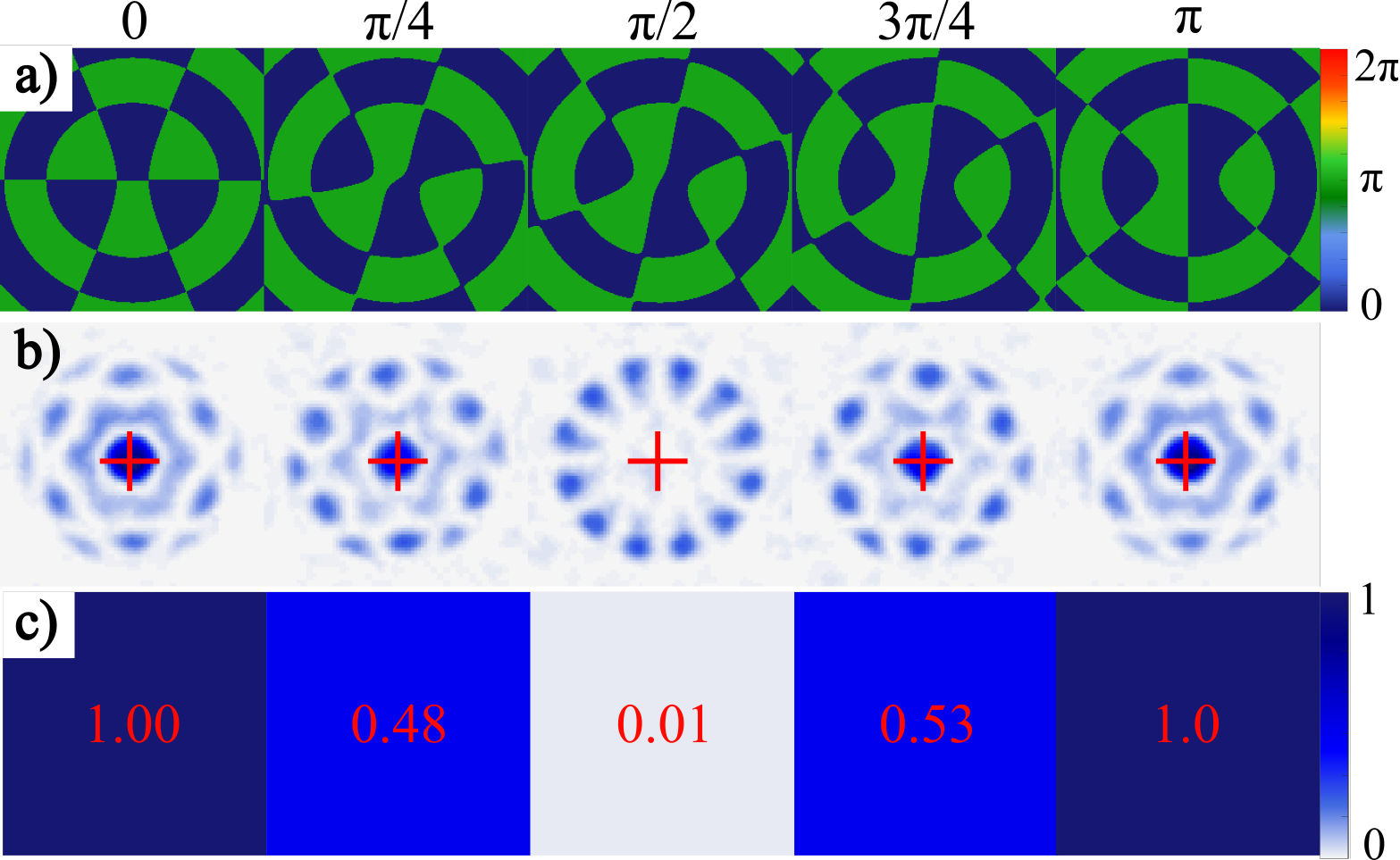}
    \caption{Far field intensity and the corresponding phase distribution on SLM2 for different $\theta_2$ values.}
    \label{measurements}
\end{figure}

As a first step, we perform a simultaneous local correlation measurement on both DOFs, polarisation and spatial mode \cite{McLaren2015}. The polarisation DOF is measured by sending the vector beam through a polarisation filter, implemented with a linear polariser oriented at different polarisation angles. The spatial DOF is measured by modal decomposition using a series of modal filters (digital holograms) encoded on a second spatial light modulator (SLM2) \cite{SPIEbook}. Here, it is important to note that since SLMs are polarisation-dependent the diffracted light in the first diffraction order is essentially polarised in the horizontal direction, acting as a horizontal linear polariser. Hence, to measure the polarisation DOF, we can in principle rotate the SLM, an almost impossible task, or alternatively insert a HWP before the SLM and rotate this instead, a more practical approach. In regards to the spatial DOF, the input vector field is projected into a series of phase holograms given by $u_m^+(\vec{r};q')+\exp(i2\theta_2)u_m^-(\vec{r};q')$, for $\theta_2\in[0,\pi]$ rad. For the sake of clarity, Fig.\ \ref{measurements} (a) shows exemplary images of the projecting phases for $\theta_2=0, \pi/4, \pi/2, 3\pi/4$ and $\pi$. Afterwards, we measure the on-axis intensity in the far field, achieved in the focal plane of a third lens (L3). Figure \ref{measurements} (b) shows example images of the far field intensity for the different values of $\theta_2$ given above and for the specific case of $\theta_1=0$, with the on-axis values marked with a red cross. The peak on-axis intensity values of all images are normalised to the highest, and stored for processing. For the sake of clarity, Fig. \ref{measurements} (c) shows a table with such values whereby, a maximum value of 1.00 was obtained for $\theta_2=0$ and $\theta_2=\pi$, intermediate values of 0.48 and 0.53 were obtained for $\theta_2=\pi/4$ and $\theta_2=3\pi/4$, respectively, and a minimum value of 0.01 was obtained for $\theta_2=\pi/2$. 

\begin{figure*}[t]
    \centering
    \includegraphics[width=1.0\textwidth]{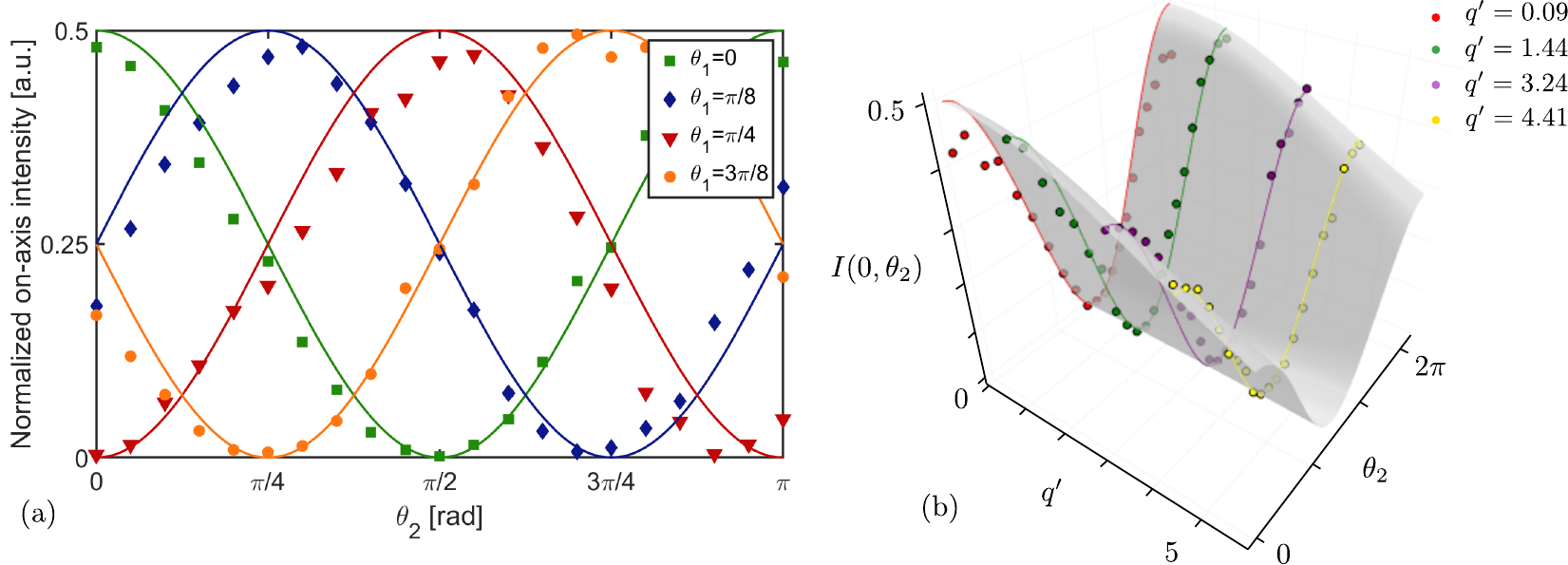}
    \caption{(a) Bell--type curve from an input helical Mathieu--Gauss vector mode with $m=3$ and $q=q'=0.09$.  The solid lines represent the theoretical predictions, and the scattered points are the experimental intensities as a function of $\theta_2$.  (b) Bell--type surface as a function of $q'$ and the orientation angle $\theta_2$ displayed on the hologram on SLM2.  The grey surface was computed through Eq.\ \ref{eq:Intensity}, whereas the coloured dots were experimentally measured. The input HMGV beam was generated with $m=3$ and $q=0.09$.}
    \label{fig:BellInequalities}
\end{figure*}

Figure \ref{fig:BellInequalities} (a), shows the normalised intensity as a function of the projection angle $\theta_2$ (the hologram displayed on the SLM) for three configurations of the polarisation angle $\theta_1$, namely, $\theta_1=0, \pi/8, \pi/4,3\pi/8$ and for the specific case $q=q'=0.09$ in Eq. \ref{eq:Intensity}.  From this data and using Eqs.\ \ref{BellParameter} and \ref{eq:E}, the calculation of the Bell parameter is straight forward and for our particular example yields the value $S=2.88\pm0.12$.  Importantly, as predicted by Eq. \ref{eq:Intensity}, the maximum value of the intensity $I(\theta_1,\theta_2)$ is a function of $q$ and $q'$. More precisely, it decreases directly proportional to the difference between $q$ and $q'$ by the factor $Q(m,q,q')$. By way of example, we plotted the intensity given by Eq. \ref{eq:Intensity} for the specific case $\theta_1=0$ and $q=0.09$, for $q'\in[0, 6]$, which is shown in Fig. \ref{fig:BellInequalities} (b). We corroborated this theoretical prediction experimentally for the specific cases $q'=0.09,1.44,3.24$ and 4.41, which are also shown in \ref{fig:BellInequalities} (b) as curves with different colours, overlapping the theoretical curve.

\begin{figure}[tb]
    \centering
    \includegraphics[width=0.49\textwidth]{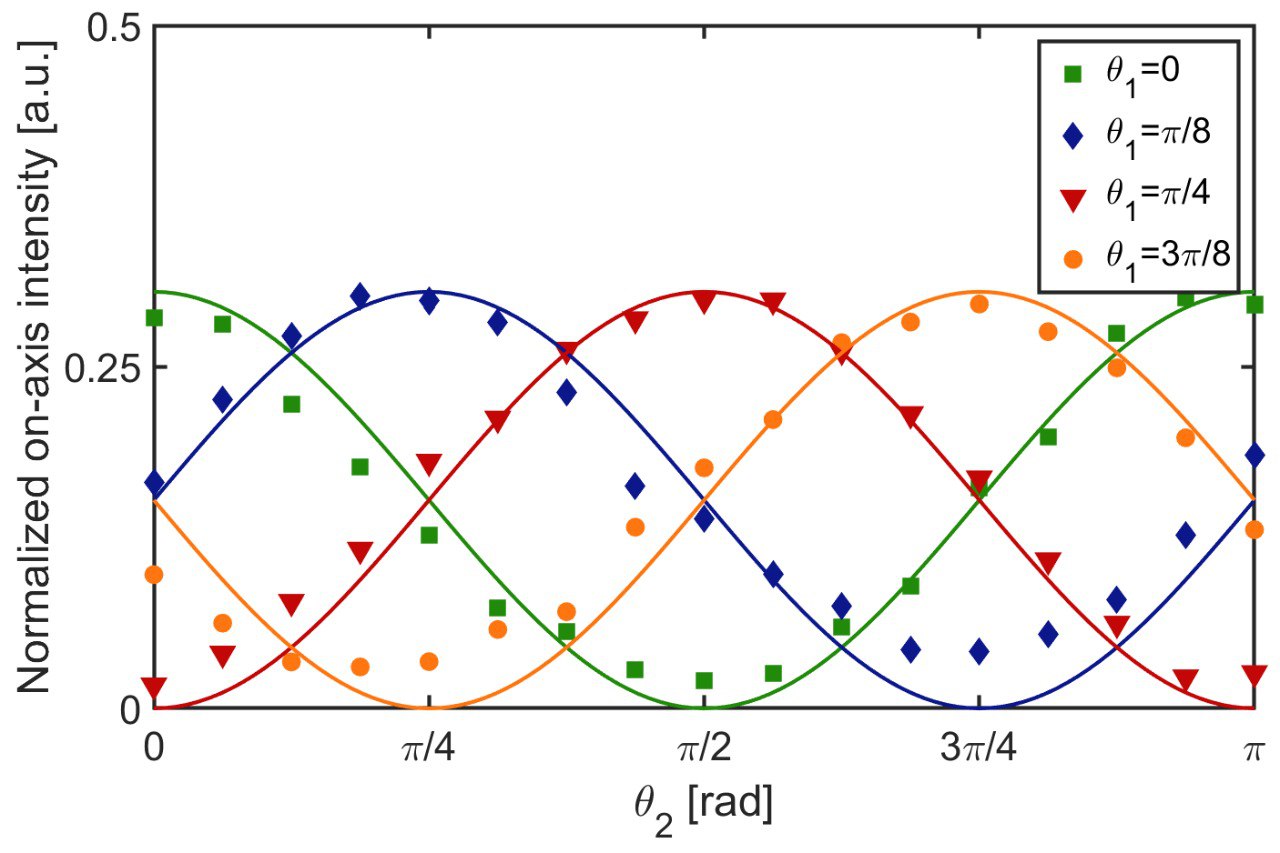}
    \caption{Bell--type curve from an input helical Mathieu--Gauss vector mode with $m=3$ and $q=0.09$ $q'=5.76$.  The solid lines represent the theoretical predictions, and the scattered points are the experimental intensities as a function of $\theta_2$.}
 \label{Bell2}
\end{figure}

Crucially, such decrease in intensity does not affect the measurement of the Bell parameter, as already mentioned above and observed by direct inspection of Eq. \ref{eq:E}, which evinces that the factor $Q(m,q,q')$ is common to the numerator and the denominator and therefore cancels out. To corroborate this experimentally, we determined the Bell parameter $S$ for the specific case $q=0.09$ and $q'=5.76$. The curves of intensity as a function of $\theta_2$, from which $S$ can be determined are shown in Fig. \ref{Bell2}. Notice that the maximum intensity value decreased significantly, but the shape of the curves remains identical. Computation of $S$ from Eq. \ref{BellParameter} yields the value $S=2.51 \pm 0.23$, which again indicates a violation of the CHSH-Bell inequality. One of the main messages we can extract from this result is that the CHSH-Bell inequality is also violated by Helical Mathieu-Gauss vector beams, but more importantly, such violation can be determined regardless of the ellipticity of the projection modes, given by the parameter $q$. A clear limitation will be the sensitivity of the specific photodetector, since the maximum and minimum intensity difference dramatically reduces as the difference between $q$ and $q'$ increases.




\section{Conclusions}
In summary, in this paper we demonstrated that Helical Mathieu-Gauss Vector (HMGV) modes, characterised by a elliptical symmetry, are locally correlated (entangled) in the classical sense, even though the distinguishing property of entangled systems, non-locality, is absent. We demonstrated this by deriving a general expression, in terms of the ellipticity of the HMGV modes that allows to measure violations to the Bell inequality for optical systems, the Clauser-Horn-Shimony-Holt-Bell (CHSH-Bell) inequality. In this context, the Bell $S$ parameter has an upper bound $|S|\leq2$ but in our experiments we measure values of $S\approx2.88$, which shows a clear violation to the Bell parameter. In the process we realised that $S$ is independent of the ellipticity of the projection mode, which implies that the violation of the same can be observed even when the ellipticity of the measured mode is not the same as the projecting mode. Finally, our demonstration that HMGV modes also violate Bell inequalities, generalises the concept of classical entanglement to vector modes with arbitrary symmetry. 

\section*{Data availability statement}
The data that support the findings of this study are available upon request from the authors.

\section*{Disclosures}
The authors declare that there are no conflicts of interest related to this article.  

\appendix

\section{Mathieu functions}\label{sec:appA}

The theory of Mathieu functions has been extensively studied in the past \cite{kirkpatrick1960tables, Gutierrez-Vega2003, meixner2006mathieu}. For instance, McLachlan \cite{arscott1968theory} treats the theory and applications of such functions in great detail. Even so, we will briefly outline some basic concepts of Mathieu functions. Let us start by writing Helmholtz equation
\begin{align}\label{eq:pde}
    \nabla_\perp^2 U + k^2 U = 0,
\end{align}
where $\nabla_\perp^2$ is the transverse Laplacian and $k$ is some constant.  In elliptical coordinates, a solution to Eq.\ \ref{eq:pde} of the form $U=R(\xi)\Phi(\eta)$ satisfies the differential equations \cite{Mathieu1868}
\begin{align} \label{eq:ang}
    \dd{\Phi}{\eta}{2} + (a - 2q\cos 2\eta)\Phi = 0,\\\label{eq:rad}
    \dd{R}{\xi}{2} - (a - 2q\cosh 2\xi)R = 0,
\end{align}
known as the angular and radial Mathieu equations, respectively.  The parameter $q$ is a dimensionless parameter and $a$ is a separation constant.  The variable $\eta\in[0,2\pi)$ is analogous to the angular coordinate, and $\xi\in[0,\infty)$ to the radial coordinate.  Solutions to Eqs.\ \ref{eq:ang} and \ref{eq:rad} are known as the angular and radial Mathieu functions, respectively.  

Angular Mathieu functions can be expanded in Fourier series as follows \cite{NIST2010}
\begin{align}
    \ce_{2n}(\eta,q) &= \sum_{j=0}^{\infty} A_{2j}^{2n}(q)\cos[2j\eta],\\
    \ce_{2n+1}(\eta,q) &= \sum_{j=0}^{\infty} A_{2j+1}^{2n+1}(q)\cos[(2j+1)\eta],\\
    \se_{2n+1}(\eta,q) &= \sum_{j=0}^{\infty} B_{2j+1}^{2n+1}(q)\cos[(2j+1)\eta],\\
    \se_{2n+2}(\eta,q) &= \sum_{j=0}^{\infty} B_{2j+2}^{2n+2}(q)\cos[(2j+2)\eta],
\end{align}
where $A_j^m$ and $B_j^m$ are the Fourier coefficients and can be computed via recurrence relations \cite{NIST2010}.  It is worth mentioning that Mathieu functions and their coefficients are readily available, for example, in the GSL numerical library \cite{GSL} and the SciPy package \cite{2020SciPy-NMeth}. A change of variable $\eta=i\xi$ allows us to obtain the Radial Mathieu functions $\Je_m(\xi,q)$ (even) and $\Jo_m(\xi,q)$ (odd). The orthogonality rules for the angular Mathieu functions are
\begin{align}
    \int_0^{2\pi} \ce_m (\eta,q) \ce_n (\eta,q) d\eta &= \pi\delta_{m,n},\\
    \int_0^{2\pi} \se_m (\eta,q) \se_n (\eta,q) d\eta &= \pi\delta_{m,n},\\
    \int_0^{2\pi} \ce_m (\eta,q) \se_n (\eta,q) d\eta &= 0.
\end{align}
Importantly, we can explore the cases with different $q$ parameter, namely
\begin{align}\nonumber
    &\int_0^{2\pi} \ce_m (\eta,q') \se_m (\eta,q) d\eta =\\
    &\hspace{0pt}\begin{cases}\nonumber
         \int_0^{2\pi}\sum_{j,j'}^{\infty} A_{2j'}^{2n}(q') B_{2j+2}^{2n+2} (q) \cos(2j'\eta) \sin[(2j+2)\eta] d\eta,\\
         \int_0^{2\pi}\sum_{j,j'}^{\infty} A_{2j'+1}^{2n+1}(q') B_{2j+1}^{2n+1}(q)\cos[(2j'+1)\eta] \sin[(2j+1)\eta] d\eta,
    \end{cases}\\
    &=\hspace{0pt}\begin{cases}\nonumber
         \sum_{j,j'}^{\infty} A_{2j'}^{2n}(q') B_{2j+2}^{2n+2} (q) \int_0^{2\pi} \cos(2j'\eta) \sin[(2j+2)\eta] d\eta,\\
         \sum_{j,j'}^{\infty} A_{2j'+1}^{2n+1}(q') B_{2j+1}^{2n+1}(q) \int_0^{2\pi} \cos[(2j'+1)\eta] \sin[(2j+1)\eta] d\eta,
    \end{cases}\\
    &=0.
\end{align}
Moreover
\begin{align}\nonumber
    &\int_0^{2\pi} \ce_m (\eta,q') \ce_m (\eta,q) d\eta =\\
    &\hspace{0pt}\begin{cases}\nonumber
         \int_0^{2\pi}\sum_{j,j'}^{\infty} A_{2j'}^{2n}(q') A_{2j}^{2n} (q) \cos(2j'\eta) \cos(2j\eta) d\eta,\\
         \int_0^{2\pi}\sum_{j,j'}^{\infty} A_{2j'+1}^{2n+1}(q') A_{2j+1}^{2n+1}(q)\cos[(2j'+1)\eta] \cos[(2j+1)\eta] d\eta,
    \end{cases}\\ \label{eq:innerce}
    &=\begin{cases}
         \pi\sum_{j=0}^{\infty} A_{2j}^{2n}(q') A_{2j}^{2n}(q), \text{ for $m=2n$},\\
         \pi\sum_{j=0}^{\infty} A_{2j+1}^{2n+1}(q') A_{2j+1}^{2n+1}(q), \text{ for $m=2n+1$}.
    \end{cases}
\end{align}
Similarly, for odd functions
\begin{align}\nonumber
    &\int_0^{2\pi} \se_m (\eta,q') \se_m (\eta,q) d\eta =\\ \label{eq:innerse}
    &\begin{cases}
         \pi\sum_{j=0}^{\infty} B_{2j+1}^{2n+1}(q') B_{2j+1}^{2n+1}(q), \text{ for $m=2n+1$},\\
         \pi\sum_{j=0}^{\infty} B_{2j+2}^{2n+2}(q') B_{2j+2}^{2n+2}(q), \text{ for $m=2n+2$}.
    \end{cases}
\end{align}

\section*{References}
\bibliographystyle{iopart-num}
\providecommand{\newblock}{}

\end{document}